\documentclass[prb,twocolumn,aps,showpacs]{revtex4}
\usepackage{times,amsmath,graphicx,latexsym}
\begin{document}

\title{Anisotropic Kosterlitz-Thouless Transition Induced by Hard-Wall Boundaries}

\author{Gary A. Williams},

\affiliation{Department of Physics and Astronomy, University of California, Los Angeles, CA 90095 USA}
\date{\today}

\begin{abstract}
The spatial dependence of the superfluid density is calculated for the Kosterlitz-Thouless transition in the presence of hard-wall boundaries, for the case of a single wall bounding the half-infinite plane, and for a superfluid strip bounded by two walls.  The boundaries induce additional vortices that cause the superfluid density to become anisotropic, with the tensor component perpendicular to the wall falling to zero at the wall, whereas the component parallel to the wall remains finite.  The effects of the boundaries are found to extend over all measured length scales, since the correlation length is infinite in the superfluid phase.
\end{abstract}

\maketitle
The behavior of a superfluid near a hard wall is still not well understood, even though many experiments are conducted in the presence of container walls.  The boundary condition at the wall is that the superflow normal to the wall should fall to zero.  There have been many guesses that this can be satisfied by having the superfluid density itself fall to zero right at the wall, but this has never been verified except perhaps in the simplest Landau-Ginsburg models.\cite{lg} Dirichlet boundary conditions at a wall are often imposed on the order parameter in perturbative renormalization expansions,\cite{dohm} but without any real justification, and since the superfluid density is not simply proportional to the order parameter in these theories its variation near the wall is often left undetermined.

Here we calculate directly the superfluid density near a hard wall for the case of the two-dimensional Kosterlitz-Thouless (KT) transition,\cite{kt} by taking into account the image vortex pairs necessary to satisfy the boundary condition.  This leads to a lowered energy of the pairs in the superfluid near the wall, and a consequent increase in the density of pairs.  The results show that the superfluid density does indeed fall to zero at the wall, but only for the component measured in the direction perpendicular to the wall, while the component parallel to the wall remains finite.  The presence of the wall thus causes the superfluid density to become an anisotropic tensor, and this perturbation is found to extend over all length scales from the wall that are being measured.  

We also consider the related problem  of two parallel walls bounding a long superfluid strip, a geometry often used in experimental measurements.  We note that several previous studies of the KT transition near hard-wall boundaries have been carried out,\cite{holz} but none of these have incorporated the anisotropy of the superfluid density, which we find to be a major effect.

These results should be of interest to experimental studies of the KT transition.  Although superfluid $^4$He films often completely wet their substrates and thus have no boundaries, it is now possible to micro-machine channels in silicon to such a small thickness that near the bulk lambda transition the helium in the channels undergoes a two-dimensional KT transition,\cite{gasparini} and the geometry and dimensions of the bounding walls can easily be varied.  The superfluid density can be measured with a small probing flow field in a given direction at frequency $\omega$, and the length scale L over which the measurement is made is then the vortex diffusion length,\cite{ahns} which is thought to vary as $\omega^{-1/2}$.  For helium films at kHz frequencies this length is known to be 4-5 orders of magnitude larger than the vortex core size.\cite{reppy}

Consider a vortex pair whose center is a distance $z$ from a wall which bounds a semi-infinite half plane of superfluid film.  The pair have a separation $r$ and the line joining their cores makes an angle $\theta$ with respect to the perpendicular to the wall.  For convenience we we will scale all of these lengths in the problem in units of the vortex core radius $a_0$.  Since the wall can be replaced by the opposite-sign image vortices, the energy of the pair, scaled by $k_B T$, is easily found to be \cite{lamb}
\begin{equation}
\frac{{U(r,\theta ,z)}}{{k_B T}} = 2\pi K_0 \left[ {\ln r + \frac{1}{2}\ln \left( {\frac{{4z^2  - r^2 \sin ^2 \theta }}{{4z^2  + r^2 \cos ^2 \theta }}} \right)} \right] + 2E_c 
\end{equation}
where $K_0  = \hbar ^2 \sigma _s^0 /m^2 k_B T$ is the bare areal superfluid density in dimensionless form, and  $E_c$ is the core energy of a vortex.  Since the energy is a function of the angle with the wall, the distribution of the thermally excited vortex pairs will be angle-dependent, and this causes the superfluid density to become anisotropic, with tensor components $K_\parallel$ and $K_\bot$.  By following the arguments of Machta and Guyer \cite{mg} for the anisotropic KT transition, it can be shown that the scaling relations for the components of the superfluid density are given by
\begin{figure}[t]
\begin{center}
\includegraphics[width=0.48\textwidth]{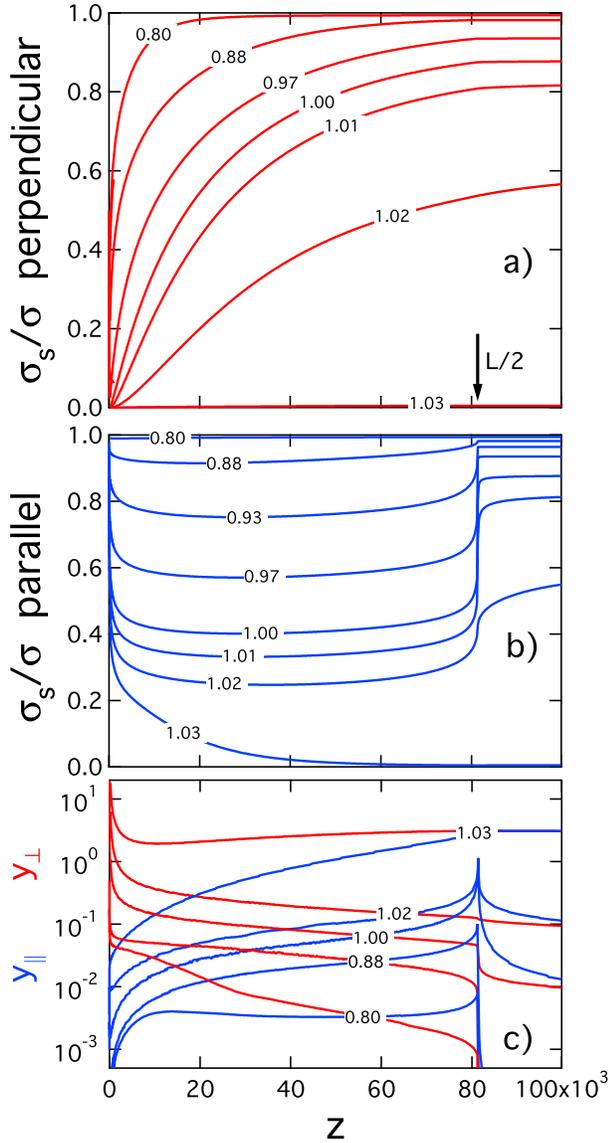}
\end{center}
\caption{Superfluid density components in the half plane a) perpendicular and b) parallel to the wall, and c) the corresponding fugacities, as a function of distance z from the wall.  The curves are labeled by their values of $T / T_{KT}$.}
\label{fig1}\end{figure}
\begin{equation}
\frac{{\partial K_\parallel  }}{{\partial \ell }} =  - 4\pi ^3 K_\parallel  \sqrt {K_\parallel  K_ \bot  } \;y_\parallel ^2 
\end{equation}
\begin{equation}
\frac{{\partial K_ \bot  }}{{\partial \ell }} =  - 4\pi ^3 K_ \bot  \sqrt {K_\parallel  K_ \bot  } \;y_ \bot ^2
\end{equation}
\begin{equation}
\frac{{\partial y}}{{\partial \ell }} = (2 - \pi \sqrt {K_\parallel  K_ \bot  } )\;y
\end{equation}
\begin{equation}
y_{\parallel ,\, \bot }^2  = y^2 \frac{4}{{\pi}}\int_{\theta _{\min } }^{\pi /2} {f(\theta )} \;\,\left( {\frac{{4z^2  - e^{2\ell } \sin ^2 \theta }}{{4z^2  + e^{2\ell } \cos ^2 \theta }}} \right)^{ - \pi \sqrt {K_\parallel  K_ \bot  } } d\theta 
\end{equation}
where $\ell  = \ln r$, $y$ is the vortex fugacity, and $f(\theta)$ equals $\cos ^2 \theta $ for the parallel moment of the fugacity $y_\parallel$ and $\sin ^2 \theta $ for  $y_\bot$.   The lower limit in the integration of Eq. 5 arises when a pair near the wall is at the angle $\theta _{\min }$ where one of the pair approaches the wall to within a core radius, and from the geometry this minimum value is   
$\theta _{\min }  = \arccos ((2z - 2)/e^{\ell } )$ when $2z - 2 < e^{\ell }$, and $\theta _{\min }  = 0$ for 
$2z - 2 \geq e^{\ell }$.  The density of pairs $n(r, z) $with separation between $r$ and $r+dr$ at the distance z is given by 
$n(r, z) = (y_\parallel ^2  + y_ \bot ^2 )/r^2$, and in the limit of large z this reduces to the infinite-plane result $y^2/r^2$.

The recursion relations are iterated using variable-step Runge-Kutta techniques, as is the integration of Eq. 5.  The iterations start at the bare scale $\ell = 0$ and terminate at the scale over which the superfluid density is measured, $\ell_m = \ln(L)$.  The initial values at $\ell = 0$ are 
$K_\parallel^0 = K_\bot^0 = K_0$, and spin-wave effects are neglected by assuming 
$\sigma _s^0 = \sigma$, the liquid density.  The initial fugacities are $y_\parallel = y_\bot = y = exp(-E_c)$, where we assume the Villain form $E_c = \pi^2 K_0/2$.  The critical value of $K_0$ at $T_{KT}$ is then $K_{0c}$ = 0.747583, and the scaled temperature is given by $T / T_{KT} =  K_{0c} / K_0$.

Figure 1 shows the components of the superfluid density and fugacity as a function of the distance $z$ from the wall for the measuring scale  $\ell_m$=12 ($L$=1.63$\times10^5$).  The superfluid density component perpendicular to the wall is zero within a core radius of the wall, and then increases smoothly to the infinite-plane value for $z > L/2$.  The perpendicular fugacity component $y_\bot$ is a maximum at the wall, reflecting the increased density of pairs parallel to the wall that drives the superfluid density to zero, and then decreases smoothly with $z$.  For distances larger than $L/2$ it then decreases to the infinite-plane value.  At high temperatures and very close to the wall $y_\bot$ can become larger than one, a high-density regime where the recursion relations are not valid, but over most of the regime where the superfluid density varies near the wall $y_\bot$ is less than one and the results should be accurate.  In Figure 1 both components of the superfluid density vary rapidly near $T_{KT}$, but there is not a sharp jump to zero because of finite-size broadening due to the cutoff of the recursion relations at the distance $L$.  This is well known \cite{barber} to give a temperature broadening above $T_{KT}$ in the form 
$\Delta T \approx (b/\ln L)^2$, where for our parameters the value of the constant $b$ is about 1.7.  

If the distance over which the perpendicular superfluid density rises to one-half of its ultimate value at large $z$ is denoted as 
$\xi _{1/2}$, it can be seen in Figure 1 that this quantity varies rapidly with temperature.  Fits to $\xi _{1/2}$ over the range $T / T_{KT}$ between 0.75 and 1.0 are best characterized by the form
$\xi _{1/2}  = 0.15 L\exp (13.3/(1-T/T_{KT}))$.  This differs from the form $\exp (b/(1-T/T_{KT})^{1/2} )$ that has been suggested as a "superfluid" coherence length for the infinite-plane case,\cite{ahns} but the reason for the difference in exponents is not clear; the precise form will require an analytic solution of the recursion relations of Eqs.\,2-5.

The component of the superfluid density parallel to the wall remains finite at the wall, but with a rapid drop over 10-40 core radii depending on the temperature.  This decrease is due to the point in the recursion relations where the pair separation is nearly 2$z$
and one of the pairs approaches to within a core radius of the wall, effectively canceling with the image in the wall.  This reduces the energy of the pair considerably, resulting in an increased vortex density that lowers the superfluid density.  The phase space of such pairs is quite limited, however, since larger pairs are cut off at $\theta _{\min }$, and the resulting fugacity component 
$y_\parallel$ is relatively small right at the wall, has a tiny peak near 10 core radii for the higher temperatures (barely visible in Fig.\,1), and then increases slowly with $z$ as the phase space increases.  For intermediate values of $z$ the parallel superfluid density is relatively constant, but then as $z$ approaches $L/2$ it increases sharply as the pair energy is reduced only for very small angles $\theta$, and finally is not reduced at all past $L/2$ where the superfluid density quickly approaches the infinite-plane value.  $y_\parallel$ peaks just at $L/2$ where the phase space increases rapidly in conjunction with the lowered pair energy, and then past $L/2$ it decreases rapidly to the infinite-plane value of $y$.
\begin{figure}[t]
\begin{center}
\includegraphics[width=0.48\textwidth]{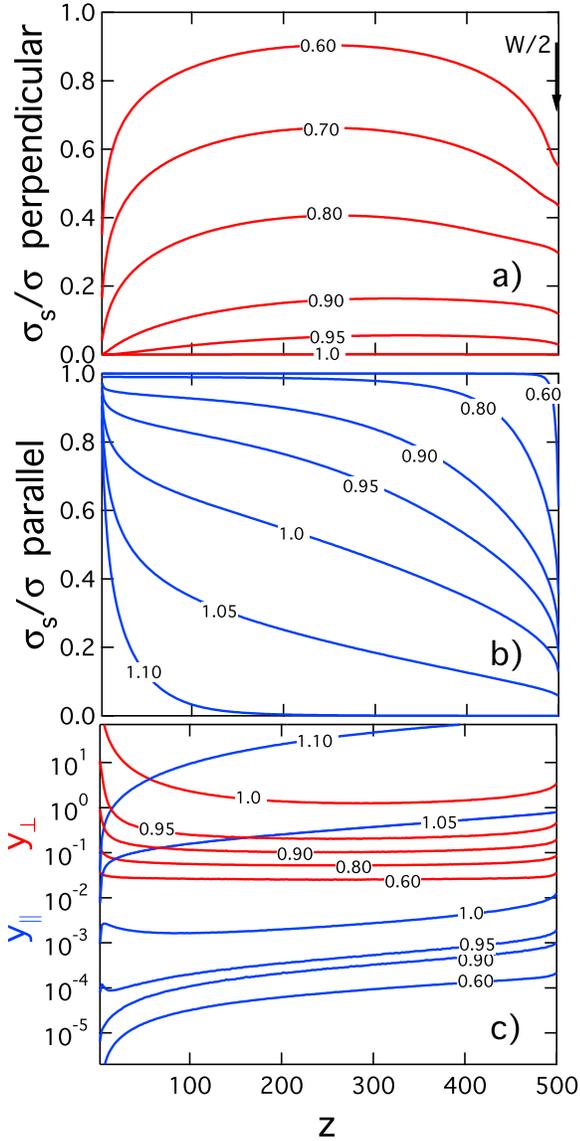}
\end{center}
\caption{Superfluid density components a) perpendicular and b) parallel to the walls of a strip of width $W$ = 1000, and c) the corresponding fugacities, as a function of distance z from the left wall.  The curves are labeled by their values of $T / T_{KT}$.}
\label{fig2}\end{figure}

\begin{figure}[t]    \centering
   \includegraphics[width=0.40\textwidth]{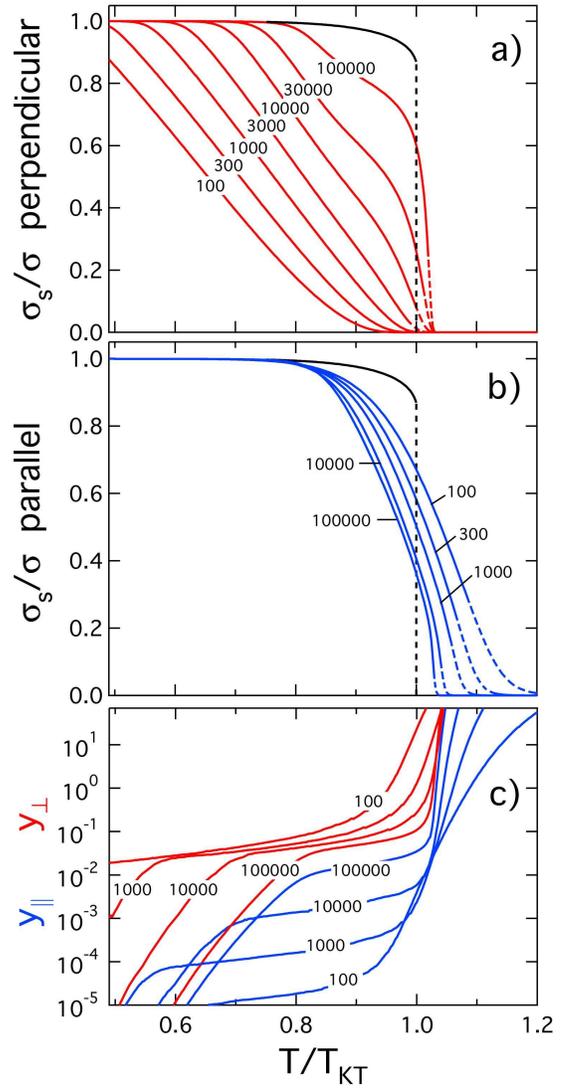} 
   \caption{Temperature dependence of the superfluid densities a) perpendicular and b) parallel to the walls of a strip of width $W$ indicated by the numbers on the curves, at a distance from the wall $z$ = $W$/4.  c) shows the corresponding fugacities, and the dashed portions of the curves indicate where the fugacities become larger than one.  The upper curve in a) and b) is the infinite-plane KT transition.}
   \label{fig:example}
\end{figure}

If these calculations are repeated at different values of $L$, the results are nearly identical when plotted as a function of $z / L$, with the only differences occurring near $T_{KT}$ due to the finite-$L$ broadening of the transition noted above.  The effect of the boundary extends over all length scales that are being measured, a consequence of the infinite Kosterlitz correlation length \cite{kt} in the superfluid phase below $T_{KT}$.  The origin of the depression of the perpendicular superfluid density at macroscopic values of $z$ can be understood by the excitation of pairs oriented parallel to the wall $(\theta = \pi /2)$, which are primarily responsible for the reduction of the perpendicular component.  It is found from Eq.\,1 that the energy to excite such pairs becomes nearly linear in the variable $z / r$ when $r$ is greater than $z$.  Hence even at very large $z$ it becomes quite favorable to excite pairs with much larger values of $r$, and these strongly reduce the perpendicular superfluid density even at temperatures well below $T_{KT}$.  It is only when the maximum value of $r$ affecting the superfluid density is limited by the measuring length $L$ that the effects of the wall are finally curtailed.

The case of two hard walls defining a superfluid strip of width W has also been investigated using the same techniques,  with the cutoff at L effectively acting as periodic boundary conditions at the ends of the strip.  The image vortices for this geometry form two infinite lines of alternating-sign vortices, and the bare energy of a pair is given by 
\begin{equation}
\begin{array}{l}
 \frac{{U(r,\theta ,z)}}{{k_B T}} = 2\pi K_0 \left[ {\ln r} \right. +  \\ 
 \\
 \left. {\frac{1}{2}\ln \left( {\frac{{(\cosh (k\beta ) - \cos (k\alpha ))\sqrt {(1 - \cos k(2z + \alpha ))(1 - \cos k(2z - \alpha ))} }}{{(\cosh (k\beta ) - \cos (2kz ))(k^2 r^2 /2)}}} \right)} \right]\\ + 2E_c  \\ 
 \end{array}
\end{equation}
where $k = \pi / W$, $\alpha = r\cos \theta$, and $\beta = r\sin \theta$.
With this potential the recursion relations are identical to Eqs.\,2-4, while in Eq.\,5 the the term in the parentheses of the integral is replaced with the argument of the second logarithm in Eq.\,6, with $r = exp(l)$.  Figure 2 shows the calculated superfluid densities and fugacities for a strip of width W = 1000, and measuring scale 
$\ell_m = 12$.  Shown are the values for $z$ between 0 and $W/2$; the results are symmetric about $W/2$.  Near $z = 0$ the characteristics are quite similar to the single-wall result, with the perpendicular superfluid density rising from zero at the wall, and the parallel component  remaining finite.  The main difference comes when $z$ approaches $W/2$.  At this point both vortices in the pair can approach the walls, greatly lowering the energy and causing a sharp dip in both components of the superfluid density even at very low temperatures.  An additional effect of the second wall is that the transition for the perpendicular component is greatly broadened out, and there is also a downward shift in the perpendicular transition temperature to well below the infinite-plane $T_{KT}$.  

When the calculation is repeated for different values of W, the curves are quite similar when plotted versus the scaled variable $z/W$, with the main differences being the broadening and $T_c$ shift of the perpendicular component.
Figure 3 shows the temperature dependence of the superfluid densities and fugacities at the point $z = W/4$, for a wide range of $W$.  The perpendicular superfluid density shows the strong downward shift of $T_c$, which appears to be a logarithmic decrease with $W$, varying as $\ln(W^{-0.05})$.  The transition is also greatly broadened, with a linear temperature decrease that seems to be independent of $W$.  In contrast, the parallel component shows about a 15\% decrease in $T_c$ that is independent of $W$, and then a $W$-dependent finite-size temperature broadening varying approximately as $(b/\ln W)^2$, similar to the single-wall broadening but with $L$ now replaced by $W$.

These main results of anisotropy extending over macroscopic length scales should also apply to superconducting films that undergo the KT transition, since strip geometries are often used in the measurements.  Finite-size effects at the KT transition are known to be important in low-$T_c$ granular films,\cite{lowtc} high-$T_c$ films,\cite{hightc,loops} and Josephson-junction arrays.\cite{lobb}  However, the superconducting case involves an additional length scale, the penetration depth $\lambda$,\cite{pierson} and to compare with experiments it will be necessary to extend the above theory to incorporate that scale.  A crude first approximation is to replace the measurement length $L$ in the above calculations with $\lambda$ if it is smaller than $L$.

It also may be possible to extend the calculation to the three-dimensional superfluid transition \cite{gwloops} by using similar techniques on the vortex loops near the wall.  This would be important for gaining a more complete understanding of the critical Casimir effect in helium films, where the boundary conditions at the upper and lower film surfaces strongly affect the magnitude of the Casimir force.\cite{dohm,loops}  The images in the wall would again lower the energy to create loops, but unfortunately analytic solutions for the energy are only known for loops
parallel to the wall,\cite{fetter} and numerical techniques will be necessary for the energies at other angles.\cite{karim}  We would expect again to find anisotropy in the superfluid density, but since in 3D the correlation length is finite this would only occur within a correlation length from the wall. 
 
Many valuable discussions with Eric Varoquaux on vortices and anisotropic superfluidity are gratefully acknowledged.  This work is supported in part by NASA, and in part by the National Science Foundation, grant DMR 05-48521.

\end{document}